\documentclass[12pt]{article}
\usepackage{epsfig,graphicx,amsmath, amssymb}

\title{Quark mass dependence of the QCD temperature transition in magnetic fields }
\author{  V.D. Orlovsky and Yu.A. Simonov \\ Institute of Theoretical and Experimental
Physics\\ 117218, Moscow, B.Cheremushkinskaya 25, Russia}

\newcommand{\be}{\begin{equation}}
\newcommand{\ee}{\end{equation}}

\def\fun#1#2{\lower3.6pt\vbox{\baselineskip0pt\lineskip.9pt
\ialign{$\mathsurround=0pt#1\hfil ##\hfil$\crcr#2\crcr\sim\crcr}}}

\newcommand{{\SD}}{\rm SD}

\newcommand{{\Mc}}{\mathcal{M}}

\newcommand{\lan}{\langle}
\newcommand{\ran}{\rangle}

\begin{document}

\maketitle
\begin{abstract}
Vacuum energy of quarks, $\varepsilon^{(q)}_{vac} = \sum_q m_q|\lan \bar q q
\ran |$ participates in the pressure balance at the temperature transition
$T_c$ and defines the dependence of $T_c$ on $m_q$. We first check this
dependence in absence of magnetic fields $eB$ vs known lattice data, and then
take into account the known strong dependence of the quark condensate on $eB$.
The resulting function $T_c(eB, m_q)$  is valid for all $eB, m_q< \sqrt\sigma$ and
explains the corresponding lattice data.
\end{abstract}
\section{Introduction}
 The QCD matter is believed to undergo a temperature transition from the low
 temperature confining state to a deconfined state of quarks and gluons. On the
 theoretical side the most detailed information on this  transition was obtained
 in numerical lattice studies, see \cite{1,2,3} for reviews. On the
 experimental side only indirect data on this transition in heavy ion
 collisions are available \cite{4,5}, however the full theory is highly needed
 both for the experiment and for astrophysical applications, e.g. in the study
 of neutron stars, see e.g. \cite{6}. Analytic studies of the QCD temperature
 transition still  predominantly based on models, which  can describe partial
 features of the transition \cite{7,8,9}, but the full analytic theory is still
 lacking.

 In 1992 one of the authors has suggested a simple but internally consistent
 mechanism of the QCD temperature transition \cite{10}, based on  a general
 nonperturbative approach in QCD, called the Field Correlator Method (FCM)
 which was formulated for   the temperature theory in \cite{11,12,13}.

 In FCM both perturbative and nonperturbative (np) dynamics are given by field
 correlators, which can be computed within the method itself \cite{14}, or taken
 from lattice data \cite{15}. Moreover, the total thermodynamic potential (free energy) of  a state includes the
 vacuum energy, which can be different in the confining and nonconfining states.

 It was exploited in \cite{10}, that the vacuum energy of the  confined state
 contains gluon  vacuum energy $\varepsilon^g_{vac}\equiv\varepsilon^g_{vac} (mag) + \varepsilon_{vac}^g
 (el)$, with the colormagnetic $\varepsilon^g_{vac}( mag)$ and colorelectric
 $\varepsilon^g_{vac} (el)$ parts, whereas in the deconfined state the
 (confining) colorelectric vacuum fields $\varepsilon^g_{vac} (el) $  are
 absent. In this way the  transition temperature $T_c$ was calculated from the
 equality of the  confined state pressure $P_I$ and the deconfined one $P_{II},
 P_I = P_{II}$, at $T=T_c$, where \be P_I = |\varepsilon_{vac}^{(g)} (el) +
 \varepsilon^{(g)}_{vac} (mag) | + P_{hadr} (T), \label{1}\ee

 \be P_{II}  = |\varepsilon_{vac}^{(g)} (mag) |+
 P_{q} (T)+ P_g(T). \label{2}\ee

 Taking for $|\varepsilon^{g}_{vac}  | $ the standard values of the gluon
 condensate \cite{16, 17} and assuming, that $|\varepsilon^{(g)}_{vac} (mag)|$
 is equal to one half of the total, as it is for zero temperature, and
 neglecting in the first approximation  $ P_{hadr} (T)$, one obtains reasonable
  values of the transition temperature $T_c$ for different values of the number
  of flavors $n_f$ as it is shown in Table I,  the upper part, in comparison with available
  lattice data.  Note, that in this calculation the input parameters are
  $|\varepsilon_{vac}^{(g)}|= 0.006$ GeV$^4$ which is a basic parameter in QCD and the
  nonperturbative interaction $ V_1(r,T)$, generating Polyakov line average, $L_f = \exp\left(-\frac{V_1(\infty, T)}{2T} \right)$,
  which is calculated from the field correlator \cite{11,12,13}. The actual parameter, entering Polyakov loops is the same, as in \cite{13}, $V_1(\infty, T_c)\cong 0.5$ GeV.

\begin{table}
\caption{Transition temperature $T_c$ for massless quarks, $n_f=0,2,3$ (the upper part), and for different nonzero $m_q$ and $n_f$ (the lower part) in comparison with lattice data.}
\begin{center}
\begin{tabular}{|c|c|c|c|c|c|}
  \hline
  $n_f$ & $m_u,$ MeV & $m_d,$ MeV & $m_s,$ MeV & $T_c,$ MeV & $T_c$ (lat), MeV \\
  \hline
  0 & - & - & - & 268 & 276 \cite{17*}  \\
  2 & 0 & 0 & - & 188 &  \\
  3 & 0 & 0 & 0 & 174 &  \\
  \hline
  2 & 3 & 5 & - & 189 & 195-213 \cite{17**} \\
  2+1 & 3 & 5 & 100 & 182 & 175 \cite{19,20} \\
  3 & 100 & 100 & 100 & 195 & 205 \cite{25}\\
  \hline
\end{tabular}
\end{center}
\end{table}

  These calculations, however,  have been done neglecting the quark vacuum
  energy,
  \be |\varepsilon_{vac}^{(q)} | = \sum^{n_f}_{i=1} m_i |\lan q_i \bar
  q_i\ran|,\label{3}\ee which is possible   for zero quark masses, and can be a good
  approximation for $n_f=2$, but not for the $(2+1)$ case,  where $m_s (2$ GeV$)
  \approx 0.1$ GeV.

  Thus for a realistic case the quark vacuum energy (3) should  be included
    in $P_I, P_{II}$, with the resulting difference $\Delta\varepsilon_{vac}^{(q)} = \varepsilon_{vac}^{(q)} (I) - \varepsilon^{(q)}_{vac} (II)
    \equiv \sum_qZ_q m_q \lan \bar q q\ran$.

    Here $Z_q$ is the factor, which takes into account the nonvanishing of $|\lan\bar q q\ran | $
     in the deconfined region as a result of a slower decrease of $|\lan \bar q q
    \ran |$ with temperature due to the finite mass $m_q$ and a finite limiting value due to the finite mass $m_q$ . Indeed, as  it was found in the lattice study of the $n_f=2+1$
    QCD \cite{19,20} the transition   temperature, obtained from strange quark  susceptibility is  10-15 MeV higher that from the  light quarks.

    In what follows we shall exploit the effective quark condensate $Z_q \lan
    \bar q  q\ran \equiv \lan \bar q q \ran_{eff}$ to distinguish from the
    standard quark condensate $\lan \bar q q \ran_{st} = (0.27$~GeV$)^3$. We shall omit below the subscript \emph{eff}, keeping the notation  $\lan \bar q q \ran_{st}$ for the standard condensate of $(0.27$~GeV$)^3$.

  In principle one can  check, whether this procedure of the
  $\varepsilon^{(q)}_{vac}$ inclusion is correct, calculating the quark mass
  dependence of  $T_c(m_q)$ for zero  m.f. and comparing it with the lattice data \cite{19,20,18}. As will be shown, indeed in both cases $T_c(m_q)$ grows with $m_q$,
  and the  concrete agreement is obtained for   reasonable values of gluon and quark condensates,
  $G_2 =0.006 $GeV$^4$,  $ |\lan \bar qq\ran | = (0.13$  GeV)$^3$.

  One important property of the vacuum quark energy is that  the quark condensate $\lan q \bar q\ran$ grows with the increasing magnetic field (m.f.) when a constant  magnetic  field  $B$
  is imposed in the system, see \cite{21} for lattice data and \cite{22} for
   analytic studies. It was shown recently in the accurate lattice studies
   \cite{21,23,25}, that $T_c(B)$ is decreasing with $B$, and the same result
   was obtained    in \cite{26,27} and in our latest work \cite{28}, where
   $\varepsilon^{(q)}_{vac}$ was not taken into account, and $T_c(B)$ was
   slowly tending to zero at large $B$.

   In the present paper, after checking the $T_c(m_q)$  dependence for zero
   m.f., but with $\varepsilon_{vac}^{(q)}$ taken into account, we calculate  $T_c(B)$
    for the $(2+1)$ case with physical quark masses and find that $T_c(B)$ is
    again decreasing with  growing $B$, but is tending to a constant limit at
    large $B$ when a reasonable (observed in $N_c=3$) dependence of $|\lan \bar
    q q \ran (B)|$ is accounted for. We also observe a quite different behavior of $T_c(B)$ in another case, when $|\lan \bar{q}q\ran|(B)$
is growing faster, than linearly, as it happens in $SU(2), \, n_f=4$ lattice data \cite{29}.

    The paper is organized as follows. In the next section we assemble together
    all formalism necessary to calculate transition temperature for nonzero
    quark masses and m.f. $B$. In section 3 we calculate $T_c(m_q)$ and compare
    with available lattice data, fixing in this way the starting $(B=0)$ quark
    vacuum energy.

    In section 4 the full derivation  of $T_c(m_q, B)$ is given for arbitrary m.f. and results of calculations are compared with lattice data.  Discussion of
    results and prospectives are given in the concluding section.

    \section{General formalism}

    We are basing the contents of this section on the results of \cite{10,11,12,13}, adding to
    that the contribution of the quark vacuum energy $\varepsilon^{(q)}_{vac}$.
    Thus in the confined state one has

    \be P_I= |\varepsilon_{vac}^{(g)}| +|\varepsilon_{vac}^{(q)}|+ P_{hadr}
    (T)\label{4}\ee
with $$
    \varepsilon^{(g)}_{vac}= \frac{\beta(\alpha_s)}{8\pi} \lan
    (F_{\mu\nu}^a)^2\ran \cong -\left(\frac{11}{3} N_c -\frac23 n_f\right)
    \frac{\alpha_s}{32\pi} \lan (F^a_{\mu\nu})^2\ran\equiv$$
    \be \equiv - \left( \frac{11}{3} N_c - \frac 23 n_f\right) \frac{G_2}{32} \equiv
    \varepsilon^{(g)}_{vac} (el) + \varepsilon^{(g)}_{vac} (mag),\label{5}\ee
    where $G_2 = \frac{2\alpha_s}{\pi} \lan (E^a_i)^2 + (H^a_i)^2\ran$,
 and $\varepsilon_{vac}^{(q)} $ given in (\ref{3}), while $P_{hadr}$ in absence
 of m.f. can be approximated by the free hadron gas expression,\be P_{hadr} (T)
 = \mp \sum_i\frac{T d_i}{2 \pi^2}\int^\infty_0 dk k^2 \ln (1\mp
 e^{(\mu-\varepsilon_i)/T}),\label{6}\ee
 with degeneracy factors $d_i$, and hadron energies $\varepsilon_i=
 \sqrt{k^2+m_i^2}.$ The  minus and plus signs refer to mesons and baryons
 respectively.

 Note, that Eq. (\ref{6}) does not take into  account high density and
 interaction corrections, as  well as the decay width and off-shell effects.

 In the deconfined  state   $\lan
 (E_i^a)^2\ran \equiv 0$, while the vector color electric interaction $V_1$
 produces fundamental  Polyakov loop \cite{30, 31}, $L_f=\exp \left( - \frac{V_1(\infty, T)}{2T}\right)$,
  hence the pressure is as in (\ref{2}) with
 \cite{11,12}
 \be \frac{1}{V_3 T^4} P_q (T) = \frac{4N_cn_f}{\pi^2} \sum^\infty_{n=1}
 \frac{(-)^{n+1}}{n^4}\varphi^{(n)}_q L^n_f \cosh \frac{\mu n}{T} \equiv
 p_q,\label{7}\ee
 where \be \varphi_q^{(n)} \equiv \frac{n^2m^2_q}{2T^2} K_2 \left(
 \frac{m_qn}{T}\right).\label{8}\ee

 For gluons the pressure contains adjoint Polyakov loops  $L_{adj}^n, L_{adj}=
  L_f^{9/4} =\exp \left( -\frac{9V_1(\infty)}{8T} \right)$
  \be \frac{1}{V_3 T^4} P_g (T) = \frac{2(N_c^2-1)}{\pi^2} \sum^\infty_{n=1}
 \frac{L^n_{adj}}{n^4}\equiv
 p_{gl}.\label{9}\ee

  As a result the transition temperature can be obtained from the equality $P_I
  =P_{II}$ in the form
  \be T_c(\mu) = \left(\frac{ \Delta|\varepsilon_{vac}| + \Delta P_{hadr}
  (T_c)}{p_q+p_g}  \right)^{1/4}\label{10}\ee
  and \be \Delta|\varepsilon_{vac} | = \Delta |\varepsilon^{(q)}_{vac} | +
  |\varepsilon^{(g)}_{vac} (el)|.\label{11}\ee
We now come to the exact form of the  $q\bar q$ interaction  $V_1(r)$ produced by
$np$ nonconfining correlator $D_1^E(x)$, \cite{11,12,30}, which  gives a
one-quark contribution $V_1(\infty, T)$,  \be V_1 (r,T) = \int^{1/T}_0 d \nu
(1-\nu T) \int^r_0 \xi d\xi D_1^E (\sqrt{\xi^2+ \nu^2}).\label{12}\ee

It was argued in \cite{12,30}, that $V_1(\infty, T)$ at $T$ around $T_c$ can be
approximated by the  formula \be V_1 (\infty, T) =\frac{0.175{~\rm
GeV}}{1.35\left( \frac{T}{T_c}\right)-1}, ~~ V_1 (\infty, T_c) \approx 0.5
~{\rm GeV}.\label{13}\ee

The  form (\ref{13}) agrees approximately with the known lattice data \cite{
35,36}. In what follows we shall follow the  reasoning of our previous studies
\cite{12,30} and use $V_1(\infty) \equiv V_1 (\infty, T_c)\approx 0.5$ GeV in
the fundamental Polyakov line $L_f = \exp \left( -
\frac{V_1(\infty)}{2T}\right)$ and the adjoint line $L_{adj} = \exp
\left(-\frac{9V_1(\infty)}{8T} \right)$.

 At this point one must take into account, that the  interaction $V_1 (r,T)$ is
 able to bind the $q\bar q$ pairs into bound states, see \cite{3} for a review
 and \cite{30,31} for analytic studies. Moreover, as shown on the lattice in
 \cite{32,33,34}, the $Q\bar Q$ interaction in the unquenched case  changes
 smoothly with temperature around $T_c$, where the confining  interaction
 $V_{conf}(R,T)$ is replaced by the nonconfining $V_1 (R, T)$. Therefore  one
 can introduce the pressure of the  bound pair and triple terms $P_{bound}$ and assume that the
 difference $\Delta P_{hadr} \equiv P_{hadr}(I) - P_{bound}(II)$ is  small
 quantity near $T_c$, which can be neglected in the first approximation. Note
 also, that in the quenched case this transition from $V_{conf}$ to $V_1$ has a
 different structure,  see e.g. \cite{35}. As it is, we are not yet able to
 explain why this smooth transition of $V_{conf}$ to $V_1$ happens in the
 unquenched  case and how it leads to the resulting QCD temperature transition,
 (work in this direction is going on).

As was discussed in \cite{30,31} and known from the lattice  data, see \cite{3}
for  a review, the binding  properties of $V_1(R,T)$ are concentrated in a
narrow region of temperatures  around $T_c$, while $V_1(\infty, T)$ gives a
piece of selfenergy for each quark and   antiquark, decreasing at large $T$. In
this way the confined pairs and triples of quarks and antiquarks near $T\approx
T_c$ go over into pairs and triples connected by the interaction $V_1(r,T)$ and
isolated quarks and antiquarks with energies augmented by a constant piece $V_1
(\infty, T)$. In some sense this transition is similar to the process of
ionization of neutral gas at increasing temperature, which finally produces the
ion-electron plasma in a  smooth continuous way.

In what follows we shall consider  $\Delta P_{hadr} (T_c)$ as a small term as
compared with $|\Delta \varepsilon_{vac}|$ and shall disregard it in the  first
approximation. We now can proceed with calculation of $T_c$ and we start with
the case $\mu=0, ~~ B=0$, when one can retain the  first terms with $n=1$ in
the sums in (\ref{7}) and (\ref{9}).

\section{The quark mass dependence of the transition temperature without magnetic fields }

To check the influence of  $\varepsilon_{vac}^{(q)} \equiv \sum_q \lan \bar q
q\ran m_q$ we  first  take it into account in
 the case of zero m.f.,  comparing
$T_c(m_q, eB=0)$ and $T_c(0,eB=0)$. Using the solution of (\ref{10}) for $T_c$
in two cases, when the $|\Delta\varepsilon_{vac}|$ is
$\frac12|\varepsilon^{(g)}_{vac}|$ and $\Delta \varepsilon_{vac }=
\frac12|\varepsilon^g_{vac}|+|\Delta\varepsilon_{vac}^{(q)}|$, the
corresponding values of $T_c$ are denoted as $T_c^{(0)}$ and $T_c^{(q)}$, one
obtains (using Eq. (\ref{14}) from \cite{12})
\be
T_c^{(q)}=\frac12 \tau^{(q)}
\left(1+ \sqrt{1+\frac{\kappa}{\tau^{(q)}}}\right) \left( 1+
\frac{m^2_q}{16(T_c^{(q)})^2}\right),\label{14}
\ee
\be
T_c^{(0)}=\frac12
\tau^{(0)} \left(1+\sqrt{1+\frac{\kappa}{\tau^{(0)}}}\right),\label{15}
\ee
where $\kappa=\frac12 V_1 (\infty), \tau^{(0)} = \left(\frac{(11-\frac23 n_f)
\pi^2 G_2}{64\cdot 12 n_f}\right)^{1/4}$ and \be \tau^{(q)} =\left[
\frac{\pi^2}{12 n_f} \left(\frac{(11-\frac23 n_f)  G_2}{64} +\sum_q m_q
|\lan \bar q  q\ran|\right)\right]^{1/4}.\label{16} \ee

For $n_f=3$ and $G_2 =0.006$ GeV$^4$ one obtains $\tau ^{(0)} =123$ MeV and
$T_c^{(0)} =  168 $ MeV.

Now, taking $m_q= 0.1$ GeV for the $s$ quark and  the contribution of $u,d$
quarks with  masses $\approx 3$ and 5 MeV respectively, one obtains $
\tau^{(q)} = 135$ MeV and $T_c^{(q)} = 181 $ MeV for $|\lan \bar s s
\ran|\approx (0.16$ GeV$)^3$ and $T_c^{(q)} =192$ MeV for $|\lan \bar s s
\ran|=(0.2 $ GeV$)^3$. This $10\div 15\%$ increase of  $T_c^{(q)}$ as compared
with $T_c^{(0)} $ is in line with lattice calculations of the same quantities
in \cite{18,19,20}.

To check quantitatively the value of the effective condensate $|\lan \bar q q
\ran|$, or equivalently of the factor $Z_q =\frac{|\lan \bar q q\ran|}{(0.27
{\rm GeV})^3}$, one can use the lattice numerical data of \cite{18} for $T_c
(m_q)$ in the $n_f =3$ case, presented in Table 2 and compare with our
predictions for $|\lan \bar q q\ran | = (0.13~{\rm GeV})^3$. One can see a good
agreement, which enables as choose this value of $|\lan \bar q q\ran|$ for our
calculations for nonzero m.f. in the next section. Note also, that in the region $m_q \ge \sqrt\sigma \sim 0.4$ GeV the quark condensate may have a nontrivial dependence on $m_q$, which influences the resulting values of $T_c(m_q)$, as can be seen in Table 2.

\begin{table}
\begin{center}
  \caption{Quark mass dependence of transition temperature from Eq. (\ref{14}) with $|\lan \bar{q}q \ran| = (0.13$ GeV$)^3$ in comparison with the lattice data from \cite{18}.}
\begin{tabular}{|c|c|c|c|c|c|c|c|}
  \hline
  $m_q$, MeV & 25 & 50 & 100 & 200 & 400 & 600 & 1000 \\
  \hline
  $T_c$ (lat), MeV & 180 & 192 & 199 & 213 & 243 & 252 & 270 \\
  $T_c$, MeV & 179 & 185 & 195 & 213 & 245 & 273 & 320 \\
  \hline
\end{tabular}
\end{center}
\end{table}

 In a similar way one can include $\varepsilon_{vac}^{(q)}$ in the
transition equation for nonzero chemical potential $\mu$, neglecting the
difference $ \Delta P_{hadr} (T_c)$ as before, one has an equality (\ref{10}),
which can be rewritten as \be T_c (\mu) =\left(
\frac{\frac12|\varepsilon_{vac}^{(g)}| + |\varepsilon^{(q)}_{vac}|}{p_q(\mu) +
p_g}\right)^{1/4},\label{17}\ee where $p_q(\mu)$ according to \cite{12} is
 \be
p_q (\mu) = \frac{n_f}{\pi^2} \left[\Phi_\nu \left(
\frac{\mu-\frac{V_1(\infty)}{2}}{T}\right) +\Phi_\nu \left(-
\frac{\mu+\frac{V_1(\infty)}{2}}{T}\right)\right], \label{18}\ee where
$\nu=m_q/T$ and \be \Phi_\nu(a) = \int^\infty_0 \frac{z^4
dz}{\sqrt{z^2+\nu^2}}\frac{1}{e^{\sqrt{z^2+\nu^2}-a}+1},\label{19}\ee and $p_{gl}$
to lowest order is independent of $\mu$ and is given by \be p_{gl} = \frac{2(N^2_c
-1)}{\pi^2 }\sum^\infty_{n=1} \frac{L^n_{adj}}{n^4}.\label{20}\ee We are now in a
position to turn on the external magnetic field.

\section{Transition temperature with the quark vacuum energy in external
magnetic field}

In principle the magnetic field  influences both phases of matter: 1) the quark
vacuum energy $\varepsilon_{vac}^{(q)}$ via the quark condensate $\lan \bar q q
\ran(B)$, 2) the gluon condensate via internal quark pair creation, 3) hadron gas
pressure, 4) quark gas pressure.

We shall disregard as before the hadron gas contribution $\Delta P_{hadr}$ and
start with the quark condensate. In this section we shall exploit the  same
mechanism of the temperature transition with the full (quark plus gluon) vacuum
energy, as in the previous section, but now in the external constant magnetic
field. To this end one can write the basic equilibrium equation

\be \left| \frac12 \varepsilon_{vac}^{(g)} + \varepsilon^{(q)}_{vac} (B)
\right| = P_g^{(0)} + \sum_q P_q (B),\label{21}
\ee
where $P_{g}^{(0)}  \equiv
p_{gl} T^4$, and use the previously found $P_q(B)$ from \cite{28}, valid for
all values of $B$ and $m_q, e_q\equiv |e_q|$,

 \begin{multline}
 P_q (B)=\frac{N_ce_qBT}{\pi^2} \sum^{\infty}_{n=1} \frac{(-)^{n+1}}{n} L^n_f\left\{ m_q K_1 \left(\frac{nm_q}{T}\right)  + \right.\\
\left. + \frac{2T}{n}\left(
\frac{e_qB+m^2_q}{e_q B}\right) K_2 \left(\frac{n}{T}\sqrt{e_qB
+m^2_q}\right) - \frac{ne_qB}{12T} K_0 \left(\frac{n}{T}
\sqrt{m^2_q+e_qB}\right)\right\}.\label{22}
\end{multline}
 For large $e_q B$ one can write $\bar P_q =\sum_qP_q(B)$ in the form
\be
\bar P_q (B) \approx \frac{N_c B T L_f}{\pi^2} \sum^{n_f}_{q=1}e_q m_q K_1\left(\frac{m_q}{T}\right).\label{23}
\ee
We take into account, that $|\lan \bar s s\ran (B) | $ grows with $eB$ in the
same way as $|\lan \bar d d \ran (B)|,$ while for $|\lan \bar u u\ran |$ the
charge $e_u$ is twice as big,  so that according to \cite{22}, one can write
\be
|\lan\bar qq\ran (B) | \equiv |\lan \bar{q}q\ran(0)|\sqrt{ 1+ \left(\frac{e_q B}{M^2_q}\right)^2
}, \, M_q^2 \approx 0.27~{\rm GeV^2}.\label{25}
\ee
 \begin{figure}[h]
  \centering
  \includegraphics[width=9cm]{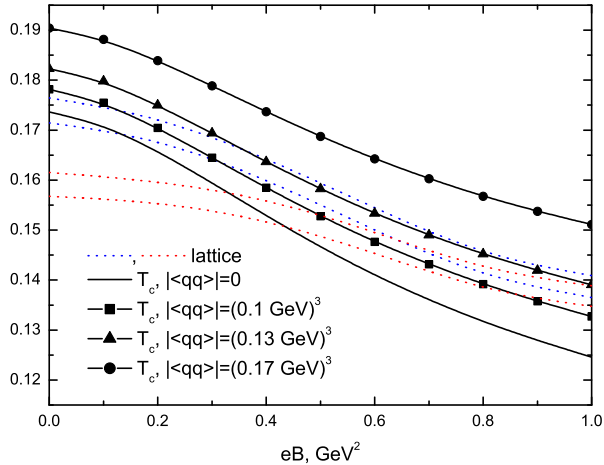}
  \caption{Transition temperature $T_c(m_q, eB, |\lan \bar{q}q\ran|_{eff})$ in GeV as a function of $eB$ for different values of the effective quark condensate, defined at zero m.f. Lattice data of \cite{23} are shown by dotted lines.}
\end{figure}
As a result the transition temperature can be found from the relation
\be
\frac{|\lan \bar{q}q\ran(0)|}{M_q^2}\sum\limits_q e_qm_q = \frac{N_cTL_f}{\pi^2} \sum\limits_q e_qm_qK_1\left( \frac{m_q}{T}\right), \label{24}
\ee
which yields $T_c(\infty) \approx 100$ MeV. In this way the asymptotic behavior
of $T_c(B)$ becomes more moderate and  its negative slope decreases at large
$eB$, again in agreement with lattice data of \cite{23}. The resulting behavior
of $T_c(B)$ with the account of quark vacuum energy is shown in Fig.~1 for
different values of $|\lan \bar q q \ran |$. Another illustration of
quark mass influence is given in Fig.~2, where a comparison is presented of our
results for the solution of Eq. (\ref{21}) and the corresponding lattice data
from \cite{21,23} for two cases: $n_f = 2+1$ with physical strange and light
quark  masses, and $n_f=3$, where all quarks have the masses of the strange
quark (lattice data from \cite{25} are given by two points at $eB =0$ and 0.8
GeV$^2$). One can see a good ($\sim 10\%)$ agreement between two sets of
results. One can see from Fig. 2, that the increasing role of
$\varepsilon^{(q)}_{vac}$ leads to the flattening of the resulting dependence
of $T_c(eB)$.

 \begin{figure}[h]
  \centering
  \includegraphics[width=9cm]{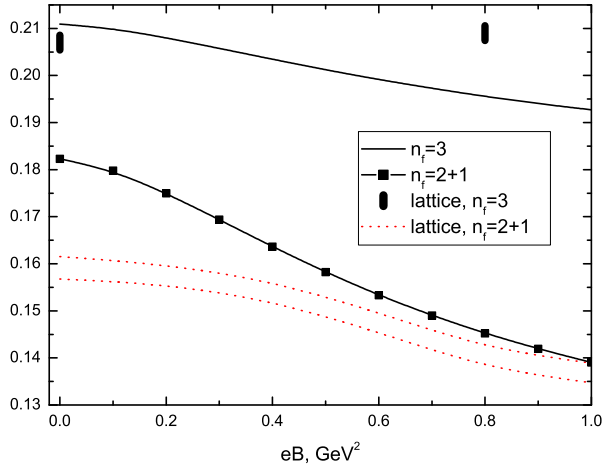}
  \caption{Transition temperature $T_c(eB)$ in GeV in two cases: $n_f=2+1$ (solid line with points) and $n_f=3$ with $m_q=0.1$ GeV and $|\lan\bar{q}q \ran|=(0.17$~GeV$)^3$ (horizontal solid line). Lattice data for $n_f=2+1$ from \cite{23} -- the area between dotted lines, and for $n_f=3$ -- two points at $eB=0$ and 0.8 GeV$^2$ from \cite{25}.}
\end{figure}

Note however, that we have neglected the  possible dependence of
$|\varepsilon_{vac}^{(g)}|$ on m.f., which should appear in higher orders of
$\alpha_s$. One can expect, that $|\varepsilon_{vac}^{(g)} (B)|$ should
decrease for growing $eB$ due to the fact, that $\alpha_s (eB)$ decreases,
since the quark loop contribution in the denominator of $\alpha_s (eB)$  in
m.f. is growing with $eB$, as shown in \cite{37}.

A decreasing behavior of $ \varepsilon_{vac}^{(g)}$ in  m.f. is obtained within
the framework of chiral perturbation theory in \cite{38}. Therefore one can expect, that the  net effect of m.f. on the $|\Delta
\varepsilon_{vac} (B)|$ would be a more mild linear growth, which implies the
partial cancellation of the effect of the quark condensate.

\section{Discussion of results and prospectives}

We have taken into account in the present paper  the effect of the quark vacuum
energy on transition temperature both with or without m.f. The case of no m.f.
helps us to fix the starting value of the strange quark condensate. Its later
growth with m.f. is predicted  by the theory, based on the chiral Lagrangian,
augmented by the   quark degrees of freedom, as well as recent lattice data.
This theory was developed before
 in \cite{22} and the resulting behavior of the quark condensate was in good
 agreement with recent accurate lattice data in \cite{21}.

 Comparing our present calculations with the previous ones in \cite{28}, one
 can  see an important role of the strange quark vacuum energy (s.q.v.e.)
 $m_s|\lan s \bar s\ran|$ at  zero m.f. and  even larger role  for  growing
 m.f. One can see in Fig. 1, that not only the asymptotics of $T_c(B)$ is
 changed by s.q.v.e. but also the slope in the region $eB\leq 1$ GeV$^2$
 becomes more flat, in better agreement with lattice data from \cite{23}.

 One more consequence of the s.q.v.e. inclusion is that the phenomenon of temperature transitions is now strongly
 dependent on the admixture of strangeness in the matter, which can be
 important for  the physics of neutron stars.
As a general outcome, one can conclude, that the main features of the quark
mass dependence of the transition temperature are accounted for  by the vacuum
quark energy $\varepsilon_{vac}^{(q)}$, and this holds both with or without
m.f.

The authors are grateful for useful discussions to M.A. Andreichikov and
B.O. Kerbikov.

\end{document}